\documentclass[epj]{svjour}

\usepackage{graphicx}
\usepackage{fancyhdr}
\def\makeheadbox{{
 }}
\def\mytitle{My title}
\def\myauthors{My name}
\def\mytype{My type of session}
\def\mysession{My session}

\def\mytitle{Lepton Flavor Violation at the LHC} 
\def\myauthors{Frank Deppisch}    
\def\mytype{Parallel Talk}
\def\mysession{Flavor Physics}

\begin{document}
\title{Lepton Flavor Violation at the LHC}
\subtitle{}
\author{Frank Deppisch\thanks{Email: frank.deppisch@manchester.ac.uk}}
\institute{Deutsches Elektronen-Synchrotron DESY, Notkestr.\ 85,
22607 Hamburg, Germany\\School of Physics and Astronomy,
University of Manchester, Manchester M13 9PL, United Kingdom}
%
\date{}
\abstract{In supersymmetric scenarios, the seesaw mechanism
involving heavy right-handed neutrinos implies sizable lepton
flavor violation (LFV) in the slepton sector. We discuss the
potential of detecting LFV processes at the LHC in mSUGRA+seesaw
scenarios and for general mixing in either the left- or
right-handed slepton sector. The results are compared with the
sensitivity of rare LFV \(\mu\to e\gamma\) decay experiments.
} 
\maketitle
\section{SUSY Seesaw Type I and Slepton Mass Matrix}\label{sec:1}
The observed neutrino oscillations imply the existence of neutrino
masses and flavor mixing, giving a hint towards physics beyond the
Standard Model. For example, the seesaw mechanism involving heavy
right handed Majorana neutrinos, which explains well the smallness
of the neutrino masses, allows for leptogenesis and induces
sizeable lepton flavor violation (LFV) in a supersymmetric
extension of the Standard Model.

If three right handed neutrino singlet fields $\nu_R$ are added to
the MSSM particle content, one gets additional terms in the
superpotential \cite{Casas:2001sr}:
\begin{equation}
\label{suppot4}
    W_\nu = -\frac{1}{2}\nu_R^{cT} M \nu_R^c + \nu_R^{cT} Y_\nu L
    \cdot H_2.
\end{equation}
Here, \(Y_\nu\) is the matrix of neutrino Yukawa couplings, $M$ is
the right handed neutrino Majorana mass matrix, and $L$ and $H_2$
denote the left handed lepton and hypercharge +1/2 Higgs doublets,
respectively. If the mass scale $M_R$ of the matrix $M$ is much
greater than the electroweak scale, and consequently much greater
than the scale of the Dirac mass matrix \(m_D=Y_\nu \langle H_2^0
\rangle\) (where \(\langle H_2^0 \rangle=v\sin\beta\) is the
appropriate Higgs v.e.v., with \(v=174\)~GeV and \(\tan\beta
=\langle H_2^0\rangle/\langle H_1^0\rangle\)), the effective left
handed neutrino mass matrix $M_\nu$ will be naturally obtained,
\begin{equation}
\label{eqn:SeeSawFormula}
    M_\nu = m_D^T M^{-1} m_D = Y_\nu^T M^{-1} Y_\nu (v \sin\beta )^2.
\end{equation}
The matrix $M_\nu$ is diagonalized by the unitary matrix
\(U_{MNS}\), yielding the three light neutrino masses:
\begin{equation}
\label{eqn:NeutrinoDiag}
    U_{MNS}^T M_\nu U_{MNS} = \textrm{diag}(m_1,m_2,m_3).
\end{equation}
The other three neutrino mass eigenstates are too heavy to be
observed directly, but, through virtual corrections, induce small
off-diagonal terms in the evolved MSSM slepton mass matrix,
\begin{eqnarray}
 m_{\tilde l}^2=\left(
    \begin{array}{cc}
        m_L^2    & (m_{LR}^{2})^\dagger \\
        m_{LR}^2 & m_R^2
    \end{array}
      \right)_{\rm MSSM}\!\!\!\!\!\!+\left(
    \begin{array}{cc}
       \delta m_L^2    & (\delta m_{LR}^{2})^\dagger \\
        \delta m_{LR}^2 & 0
    \end{array}
      \right)\!\!,
\end{eqnarray}
leading to observable LFV processes. These corrections in leading
log approximation are \cite{Hisano:1999fj}
\begin{eqnarray}
\label{left_handed_SSB2}
    \delta m_{L}^2 &=& -\frac{1}{8 \pi^2}(3m_0^2+A_0^2)(Y_\nu^\dag L Y_\nu),\\
    \delta m_{LR}^2 &=& -\frac{3 A_0 v \cos\beta}{16\pi^2}(Y_l Y_\nu^\dag L Y_\nu),
\end{eqnarray}
where $L_{ij} = \ln(M_{GUT}/M_i)\delta_{ij}$, and \(m_0\) and
\(A_0\) are the common scalar mass and trilinear coupling,
respectively, of the minimal supergravity (mSUGRA) scheme. The
product of the neutrino Yukawa couplings $Y_\nu^\dagger L Y_\nu$
entering these corrections can be determined by inverting
(\ref{eqn:SeeSawFormula}),
\begin{equation}
\label{eqn:yy}
    Y_\nu =
        \frac{1}{v\sin\beta}\textrm{diag}(\sqrt{M_i})
        \!\cdot\!R\!\cdot\!\textrm{diag}(\sqrt{m_i})\!\cdot\! U_{MNS}^\dagger,
\end{equation}
using neutrino data as input for the masses \(m_i\) and
\(U_{MNS}\), and evolving the result to the unification scale
$M_{GUT}$. The unknown complex orthogonal matrix $R$ may be
parametrized in terms of 3 complex angles $\theta_i=x_i +i y_i$.

\section{LFV Rare Decays and LHC Processes}\label{sec:2}
At the LHC, a feasible test of LFV is provided by production of
squarks and gluinos, followed by cascade decays via neutralinos
and sleptons \cite{Agashe:1999bm,Andreev:2006sd}:
\begin{eqnarray}\label{eqn:LHCProcesses}
    pp             &\to& \tilde q_\alpha \tilde q_\beta, \tilde g \tilde q_\alpha, \tilde g
    \tilde g,\nonumber\\
    \tilde q_\alpha(\tilde g)&\to& \tilde\chi^0_2 q_\alpha(g),\nonumber\\
    \tilde\chi^0_2 &\to& \tilde l_a l_i,\nonumber\\
    \tilde l_a     &\to& \tilde\chi^0_1 l_j,
\end{eqnarray}
where \(a,b,i\) run over all sparticle mass eigenstates including
antiparticles. LFV can occur in the decay of the second lightest
neutralino and/or the slepton, resulting in different lepton
flavors, \(\alpha\neq\beta\). The total cross section for the
signature \(l^+_\alpha l^-_\beta + X\) can then be written as
\begin{eqnarray}\label{eqn:LHCProcess}
   &&\sigma(pp\to l^+_\alpha l^-_\beta+X) = \nonumber\\
   && \Bigl\{
   \quad \sum_{a,b}\sigma(pp\to\tilde q_a\tilde q_b)\times Br(\tilde q_a\to\tilde\chi^0_2
         q_a)\nonumber\\
   &&\quad+\sum_{a}\sigma(pp\to\tilde q_a\tilde g)  \times Br(\tilde q_a\to\tilde\chi^0_2 q_a)\nonumber\\
   &&\quad+\quad\,\,\,\,\, \sigma(pp\to\tilde g\tilde g)\times Br(\tilde g\to\tilde\chi^0_2
        g) \nonumber\\
   && \Bigr\}\times Br(\tilde\chi^0_2\to l_\alpha^+l_\beta^-\tilde\chi^0_1),
\end{eqnarray}
where \(X\) can involve jets, leptons and LSPs produced by lepton
flavor conserving decays of squarks and glui\-nos, as well as low
energy proton remnants. The cross section is calculated at the LO
level \cite{Dawson:1983fw} with 5 active quark flavors, using
CTEQ6M PDFs \cite{Pumplin:2002vw}. Possible signatures of this
inclusive process are:
\begin{itemize}
    \item $ l_i l_j  \quad\,\,\, + 2\textrm{jets} + E_{miss}$
    \item $ l_i l_j  \quad\,\,\, + 3\textrm{jets} + E_{miss}$
    \item $ l_i l_j l_k l_k      + 2\textrm{jets} + E_{miss}$,
\end{itemize}
with at least two leptons \(l_i, l_j\) of unequal flavor.

The LFV branching ratio \(Br(\tilde\chi^0_2\to
l_\alpha^+l_\beta^-\tilde\chi^0_1)\) is for example calculated in
\cite{Bartl:2005yy} in the framework of model-independent MSSM
slepton mixing. In general, it involves a coherent summation over
all intermediate slepton states.

\begin{figure}[t]
\centering
\includegraphics[clip,width=0.49\textwidth]{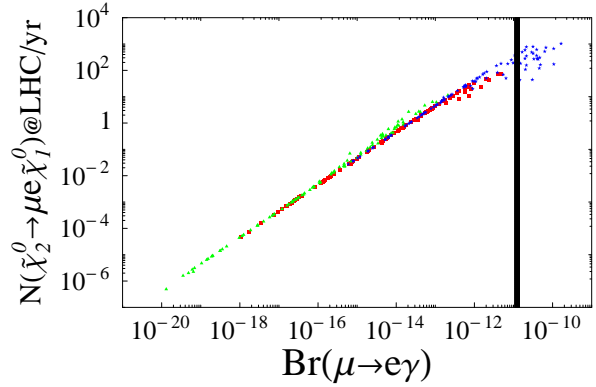}
\caption{Correlation of the number of
\(\tilde\chi_2^0\to\mu^+e^-\tilde\chi_1^0\) events per year at the
LHC and \(Br(\mu\to e\gamma)\) in mSUGRA scenario C'
(\(m_0=85\)~GeV, \(m_{1/2}=400\)~GeV, \(A_0=0\)~GeV,
\(\tan\beta=10\)~GeV, \(\textrm{sign}\mu=+\)) for the case of
hier.\ $\nu_{R/L}$ (blue stars), deg.\ $\nu_R$/hier.\ $\nu_L$ (red
boxes) and deg.\ $\nu_{R/L}$ (green triangles). The neutrino
parameters are scattered within their experimentally allowed
ranges~\cite{Maltoni:2003sr}. For degenerate heavy neutrino
masses, both hierarchical (green diamonds) and degenerate (blue
stars) light neutrino masses are considered with real $R$ and
$10^{11}\ {\rm GeV}<M_R < 10^{14.5}\ {\rm GeV}$. In the case of
hierarchical heavy and light neutrino masses (red triangles),
$x_i$ is scattered over $0<x_i <2\pi$ while $y_i$ and $M_i$ are
scattered in the ranges allowed by leptogenesis and perturbativity
\cite{Deppisch:2005rv}. An integrated LHC luminosity of
\(100\textrm{fb}^{-1}\) is assumed. The current limit on
\(Br(\mu\to e\gamma)\) is displayed by the vertical line.}
\label{fig:br12_N2}
\end{figure}

\begin{figure}[t]
\centering
\includegraphics[clip,width=0.49\textwidth]{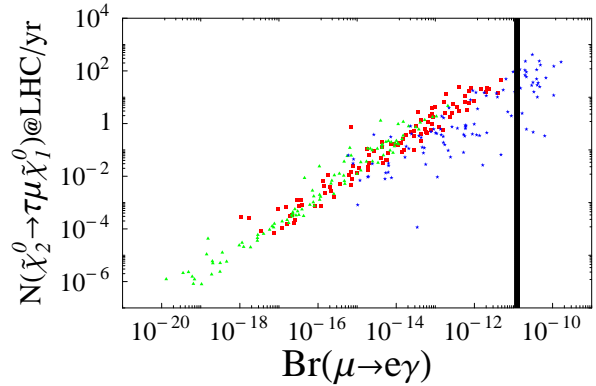}
\caption{Same as Figure~\ref{fig:br12_N2}, but correlating
\(\tilde\chi_2^0\to\tau^+\mu^-\tilde\chi_1^0\) with \(Br(\mu\to
e\gamma)\).} \label{fig:br23_N2}
\end{figure}

As a sensitivity comparison it is useful to correlate the expected
LFV event rates at the LHC with LFV rare decays (see
\cite{Deppisch:RareDecays} and references therein for a discussion
of LFV rare decays in SUSY Seesaw Type I scenarios). This is shown
in Figures~\ref{fig:br12_N2} and \ref{fig:br23_N2} for the event
rates \(N(\tilde\chi_2^0\to\mu^+e^-\tilde\chi_1^0)\) and
\(N(\tilde\chi_2^0\to\tau^+\mu^-\tilde\chi_1^0)\), respectively,
originating from the cascade reactions (\ref{eqn:LHCProcesses}).
Both are correlated with \(Br(\mu\to e\gamma)\), yielding maximum
rates of around \(10^{2-3}\) per year for an integrated luminosity
of \(100\textrm{fb}^{-1}\) in the mSUGRA scenario C'
\cite{Battaglia:2001zp}, consistent with the current limit
\(Br(\mu\) \(\to e\gamma)<10^{-11}\). The MEG experiment at PSI is
expected to reach a sensitivity of \(Br(\mu\to e\gamma)\approx
10^{-13}\).

\begin{figure}[t]
\centering
\includegraphics[clip,width=0.45\textwidth]{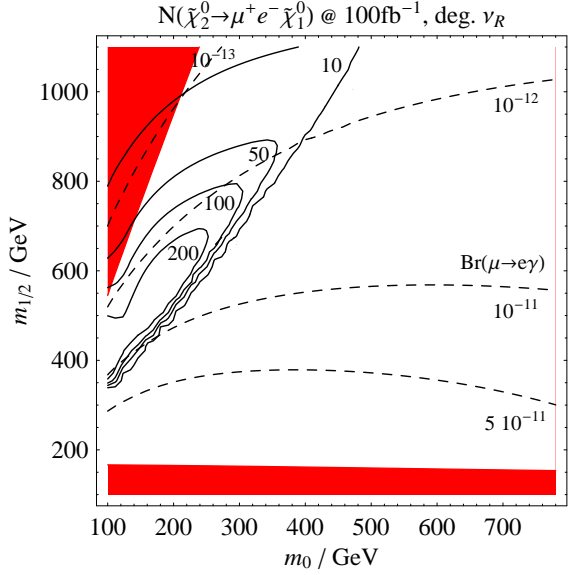}
\caption{Contours of the number of
\(\tilde\chi_2^0\to\mu^+e^-\tilde\chi_1^0\) events at the LHC with
an integrated luminosity of \(100\textrm{fb}^{-1}\) (solid) and of
\(Br(\mu\to e\gamma)\) in the \(m_0-m_{1/2}\) plane . The
remaining mSUGRA parameters are as in Figure~\ref{fig:br12_N2}.
The neutrino parameters are at their best fit values
\cite{Maltoni:2003sr}, with \(m_{\nu_1}=0\) and a degenerate r.h.
neutrino mass \(M_R=10^{14}\)~GeV. The shaded (red) areas are
already excluded by mass bounds from various experimental
sparticle searches.} \label{fig:scan_lhc_seesaw}
\end{figure}

The correlation is approximately independent of the neutrino
parameters, but highly dependent on the mSUGRA parameters. This is
contemplated further in Figure~\ref{fig:scan_lhc_seesaw},
comparing the sensitivity of the signature
\(N(\tilde\chi_2^0\to\mu^+e^-\tilde\chi_1^0)\) at the LHC with
\(Br(\mu\to e\gamma)\) in the mSUGRA \(m_0-m_{1/2}\) parameter
plane. LHC searches can be competitive to the rare decay
experiments for small \(m_0\approx200\)~GeV. Tests in the
large-\(m_0\) region are severely limited by collider kinematics.

\begin{figure}[t!]
\centering
\includegraphics[clip, width=0.45\textwidth]{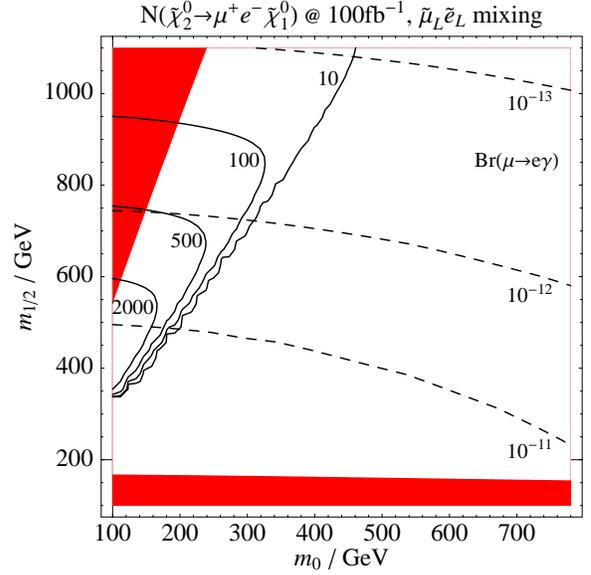}
\caption{Contours of the events per year
\(N(\tilde\chi_2^0\to\mu^+e^-\tilde\chi_1^0)\) for maximal
\(\tilde e_L\tilde\mu_L\) mixing at the LHC with an integrated
luminosity of \(100\textrm{fb}^{-1}\) in the \(m_0-m_{1/2}\) plane
(solid lines). The remaining mSUGRA parameters are:
\(A_0=-100\)~GeV, \(\tan\beta=10\), \(\textrm{sign}(\mu)=+\).
Contours of \(Br(\mu\to e\gamma)\) are shown by dashed lines. The
shaded (red) areas are forbidden by mass bounds from various
experimental sparticle searches.} \label{fig:scan_lhc maxmix_L}
\end{figure}

Up to now we have considered LFV in the class of type I SUSY
seesaw model described in Section~\ref{sec:1}, which is
representative of models of flavor mixing in the left-handed
slepton sector only. However, it is instructive to analyze general
mixing in the left- and right-handed slepton sector, independent
of any underlying model for slepton flavor violation. The easiest
way to achieve this is by assuming mixing between two flavors
only, which can be parametrized by a mixing angle \(\theta_{L/R}\)
and a mass difference \((\Delta m)_{L/R}\) between the sleptons,
in the case of left-/right-handed slepton mixing,
respectively\footnote{This is different to the approach in
\cite{Bartl:2005yy}, where the slepton mass matrix elements are
scattered randomly.}. In particular, the left-/right-handed
selectron and smuon sector is then diagonalized by
\begin{equation}\label{eqn:TwoFlavorDiag}
    \left(
    \begin{array}{c}
      \tilde l_1 \\
      \tilde l_2 \\
    \end{array}
    \right)
    = U\cdot
    \left(
    \begin{array}{c}
      \tilde   e_{L/R} \\
      \tilde \mu_{L/R} \\
    \end{array}\right)
\end{equation}
with
\begin{equation}
    U=\left(\begin{array}{cc}
      \cos\theta_{L/R}  & \sin\theta_{L/R} \\
      -\sin\theta_{L/R} & \cos\theta_{L/R} \\
    \end{array}\right),
\end{equation}
and a mass difference \(m_{\tilde l_2}-m_{\tilde l_1}=(\Delta
m)_{L/R}\) between the slepton mass eigenvalues\footnote{For
left-handed slepton mixing, \(\theta_L\) and \((\Delta m)_L\) are
also used to describe the sneutrino sector.}. The LFV branching
ratio \(Br(\tilde\chi_2^0\to\mu^+e^-\tilde\chi_1^0)\) can then be
written in terms of the mixing parameters and the flavor
conserving branching ratio \(Br(\tilde\chi_2^0\to
e^+e^-\tilde\chi_1^0)\) as
\begin{eqnarray}\label{eqn:TwoFlavorMixing}
    Br(\tilde\chi_2^0\to\mu^+e^-\tilde\chi_1^0)&=&
    2\sin^2\theta_{L/R}\cos^2\theta_{L/R} \nonumber\\
    &\times&\frac{(\Delta m)^2_{L/R}}{(\Delta m)^2_{L/R}+\Gamma^2_{\tilde
    l}} \nonumber\\
    &\times&Br(\tilde\chi_2^0\to e^+e^-\tilde\chi_1^0),
\end{eqnarray}
with the average width \(\Gamma_{\tilde l}\) of the two sleptons
involved. Maximal LFV is thus achieved by choosing
\(\theta_{L/R}=\pi/4\) and \((\Delta m)_{L/R}\gg\Gamma_{\tilde
l}\). For definiteness, we use \((\Delta m)_{L/R}\) \(=0.5\)~GeV.
The results of this calculation can be seen in
Figures~\ref{fig:scan_lhc maxmix_L} and \ref{fig:scan_lhc
maxmix_R}, which show contour plots of
\(N(\tilde\chi_2^0\to\mu^+e^-\tilde\chi_1^0)\) in the
\(m_0-m_{1/2}\) plane for maximal left- and right-handed slepton
mixing, respectively. Also displayed are the corresponding
contours of \(Br(\mu\to e\gamma)\). We see that the present bound
\(Br(\mu\to e\gamma)=10^{-11}\) still permits sizeable LFV signal
rates at the LHC. However, \(Br(\mu\to e\gamma)<10^{-13}\) would
largely exclude the observation of such an LFV signal at the LHC.
\begin{figure}[t!]
\centering
\includegraphics[clip, width=0.45\textwidth]{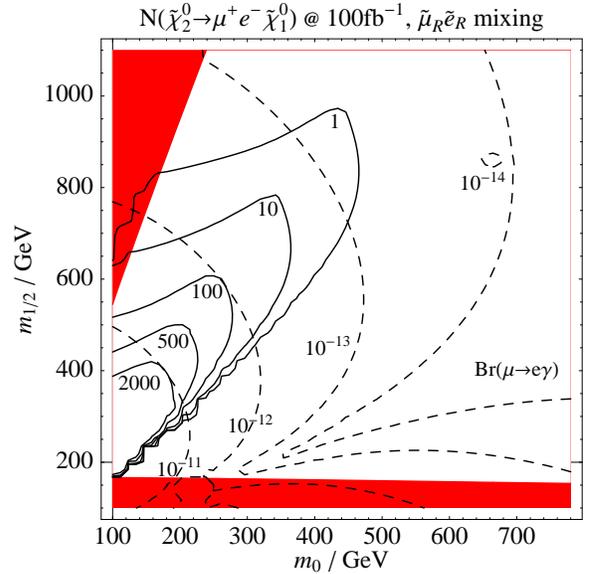}
\caption{As in Figure~\ref{fig:scan_lhc maxmix_L} but for maximal
\(\tilde e_R\tilde\mu_R\) mixing.}\label{fig:scan_lhc maxmix_R}
\end{figure}
\section*{Acknowledgments}
The author would like to thank S. Albino, D. Ghosh and R. R\"uckl
for the collaboration on which the presentation is based.

\end{document}